\documentclass[copyright,creativecommons]{eptcs}
\providecommand{\event}{2007} 
\providecommand{\volume}{??}
\providecommand{\anno}{20??}
\providecommand{\firstpage}{1}
\providecommand{\eid}{??}
\usepackage{breakurl}        
\usepackage{amssymb}
\usepackage{graphicx}
\usepackage{multirow}

\newcommand{\comp}{\mathbin{\raise 0.6ex\hbox{\oalign{\hfil$\scriptscriptstyle
     \mathrm{o}$\hfil\cr\hfil$\scriptscriptstyle\mathrm{9}$\hfil}}}}
\newenvironment{reason}{\begin{tabbing}\hspace{2em}\=\kill}{\end{tabbing}\vspace
{-2.5ex}}

\newcommand{\br}{\begin{reason}}
\newcommand{\er}{\end{reason}}
\def\z@rel#1{\mathrel{\mathstrut{#1}}}
\def\Defs{\ \triangleq \ }

\newcommand{\ccc}[3]{#1\,\,\mbox{$\lhd$}\ #2\ \mbox{$\rhd$}\ #3}
\newcommand{\pp}[1]{\mathbin{\makebox[0em]{~}_{#1}\oplus}}

\newcommand{\wP}[2]{wp.{#1}.{#2}}

\title{Quantitative Safety: Linking Proof-Based Verification with Model Checking for Probabilistic Systems}
\author{Ukachukwu Ndukwu \thanks{The author is a recipient of the Australian Commonwealth Endeavour International
Postgraduate Research Scholarship (E-IPRS) while pursuing a Ph.D. in the Information and Networked Systems Security (INSS) research group at Macquarie University, Australia.}
\institute{Department of Computing \\ Macquarie University,
NSW 2109, Sydney, Australia.}
\email{ukndukwu@science.mq.edu.au}
}
\def\titlerunning{Linking Proof-Based Verification with Model Checking}
\def\authorrunning{Ukachukwu Ndukwu}

\begin{document}
\maketitle

\begin{abstract}
This paper presents a novel approach for augmenting proof-based verification with performance-style analysis of the kind employed in state-of-the-art model checking tools for probabilistic systems.\ Quantitative safety properties usually specified as probabilistic system invariants and modeled in proof-based environments are evaluated using bounded model checking techniques \cite{CBC03}.

Our specific contributions include the statement of a theorem that is central to model checking safety properties of proof-based systems, the establishment of a procedure; and its full implementation in a prototype system (YAGA) which readily transforms a probabilistic model specified in a proof-based environment to its equivalent verifiable PRISM \cite{PRISM} model equipped with reward structures \cite{KNP07}.\ The reward structures capture the exact interpretation of the probabilistic invariants and can reveal succinct information about the model during experimental investigations.\ Finally, we demonstrate the novelty of the technique on a probabilistic library case study.


\end{abstract}

\section{Introduction}
There are two main techniques for investigating quantitative properties of {\it probabilistic} systems behaviours:

{\it Probabilistic model checking} comprises the formalisation of a model (which describes a system operationally), and a suite of algorithms to analyse various correctness and performance properties (usually expressed as a probabilistic temporal logic) of the model.

{\it Proof-based verification} on the other hand are practical applications of deductive proof methods to establish a link between the operational description of the model and the desired properties.

Over the years these two techniques have developed almost independently.\ The goal of this paper is to establish a formal link between them in a way that has never been previously explored.\ Our intention is to make such linkage beneficial to practitioners of both techniques, and hence ensure future applicability of the proposed practice especially on an industrial scale.

The B-Method \cite{Abrial96} is an industrial-strength specification language for describing large-scale abstract system behaviours.\ The method's development process ensures that specifications gradually evolve via {\it refinement} to implementable code.\ The probabilistic B (pB) \cite{Hoang05} extends the B-Method to incorporate probability.

PRISM \cite{PRISM} is a probabilistic model checker which accepts probabilistic models described in its modeling language --- a simple state-based language.
\ Three types of probabilistic models are supported directly --- Discrete Time Markov Chains (DTMCs), Markov Decision Processes (MDPs), and Continuous Time Markov Chains (CTMCs).\ This work is based on the the MDP type of the language.

One of the key features of the pB language is the statement of invariants --- they provide a means of describing properties required to maintain the integrity of a system under construction.\ In an attempt to establish a generalised view of probabilistic system invariant properties, Hoang T.S. {\it et al.} \cite{HJRMM03} originally explored the use of ``expectations'' as system invariants in proof-based systems.\ Including probabilistic invariants in systems design is a useful means of enforcing quantitative safety properties --- for example, tolerance for expected error in a probabilistic system behaviour.

Discharging a pB machine's safety proof obligation involves the verification (by a human or automated support) that indeed the expectation holds for all executions of the machine; but sometimes the prover fails to establish this goal.\ It is pertinent to mention that not all the undischargeable proof obligations are malignant to the overall performance outlook of the final machine for deployment.\ However, in safety-critical systems development, this assumption cannot be taken lightly hence the need to explore other techniques of gaining intuition into the failure of the provers to discharge their proof obligations.

For standard (non-probabilistic) B machines, a prototype system \cite{LB03} which incorporates a model checking tool has been used to detect various errors in simple machine specifications.\ However, for the probabilistic counterparts, we show how to link the model checking facility of PRISM with abstract systems specified in pB via the latter's probabilistic invariants.\ The importance of such a link is to enable pB designers explore their models experimentally; such an exploration is likely to reveal undesirable performance attributes of their models, and in particular guide them with a decision in the event that the invariants fail to hold.

Our technique is as follows.\ Given a proof-based machine specified in pB, we generate its equivalent probabilistic action systems representation (in the PRISM language) fully augmented with reward structures \cite{KNP07} inherited from the pB machine's expectations, and either confirm (or refute) the statement over the expectations.\ In the event that the statement fails to hold, we get an intuition of the possible cause(s) of failure.\ Our specific contributions are as follows:

\begin{itemize}
\item [(a)]\ The statement of a theorem that forms the implicit link between a pB invariant and its reward-based model checking equivalent.

\item [(b)]\ A procedure for transforming a pB machine specified in a proof-based environment to its equivalent PRISM representation.

\item [(c)]\ A prototype system to automate the procedure.\ In addition, we equip the resultant PRISM model with reward structures inherited from the pB machine's expectations, to allow for experiments.

\item [(d)]\ A demonstration of the novelty of our technique on a small case study of a probabilistic library system.
\end{itemize}
Overall, this paper is structured as follows.\ We introduce the pGSL and its expectations in sec. 2, and set the theoretical foundation of our procedure in secs. 3, and 4.\ The automation of that procedure is in sec. 5.\ A practical demonstration of our technique is in secs. 6 and 7.\ Finally, we conclude in sec. 8.
\begin{figure}
\begin{center}
$
\small
\begin{array}{lll}
\mbox{command} & \mbox{command} & \mbox{\quad  transformer semantics }\\
\mbox{ \ \ name}& \mbox{ \ ($comm$)}& \mbox{ \quad \quad  ($wp.{{comm}}.{E})$}\\
\hline\\

\mbox{identity} & \mbox{skip} &  \mbox{$E$}   \\
\mbox{assignment} & \mbox{$x := f$}  & \mbox{$E[x := f]$}\\
\mbox{composition} & \mbox{{$r;r^{'}$}} & \mbox{$\wP{r}{(\wP{r^{'}}{E})}$}\\
\mbox{choice} & \mbox{$\ccc{r}{G}{r^{'}}$}& \mbox{$\ccc{\wP{r}{E}}{G}{\wP{r^{'}}{E}}$} \\
\mbox{probability} & \mbox{$r \pp{p} \ r^{'}$} & \mbox{$\wP{r}{E} \ \pp{p} \ \wP{r^{'}}{E}$}\\
\mbox{nondeterminism} & \mbox{${r \ \sqcap \ r^{'}}$} & \mbox{$\wP{r}{E} \ $min$ \ \wP{r^{'}}{E}$}\\
\mbox{strong iteration} & \mbox{{do} ~$G \rightarrow {r}$ ~{od}} & \mbox{$\mu X \bullet (\ccc{\wP{r}{X}}{G}{{E}}$)}\\
\mbox{weak iteration} & \mbox{It $r$ tI} & \mbox{$\nu X \bullet (\wP{r}{X}\ $min$ \ E)$}\\
\\
\hline\\

\end{array}
$
\caption{\rm Structural definition of the expectation transformer-style semantics.} \label{fig:semantics}\vspace{-.5cm}
\end{center}
\end{figure}

\section{Probabilistic Generalised Substitution Language {\it pGSL}}
Abrial's Generalised Substitution Language {\it GSL} \cite{Abrial96} is based on Dijkstra's weakest-precondition $wp$ semantics of describing computations and their meaning \cite{Dijkstra76}.\ The semantics, expressive in the {\it B-Method} (B) \cite{Abrial96}, defines the concept of an ``abstract machine''.\ The {\it Abstract Machine Notation (AMN)} explores B's capabilities via {\it refinement} for incrementing designs  such that relevant system properties are always preserved.\ The complete framework supports the development of provably correct systems.

The logic {pGSL} \cite{Morgan98} is a smooth extension of the {GSL}, in which the standard boolean values --- representing certainty are replaced by real-values --- representing probability.\ Its logical framework is the {\it probabilistic Abstract Machine Notation (pAMN)} \cite{Hoang05}, an extension of the standard {AMN}.\ Its specification is based on the probabilistic B (pB).\ The syntactic structure of the {pGSL} is rich enough to permit the specification of abstract probabilistic system behaviours.\ Details of the GSL can be found in \cite{Abrial96},
while that of its theoretical and practical extensions are in \cite{Morgan98} and \cite{HJRMM03,Hoang05} respectively.

An important component of the {pGSL} is the specification of probabilistic invariants to ensure consistency between system designs.\ This is indeed crucial since it also assures that undesirable operation sequences do not lead to a violation of the critical properties of a system.\ In fact, the semantics of the {pGSL} which is based on the expectation transformer-style semantics of {pGCL} \cite{MM04} (shown in Fig. {\ref{fig:semantics}}) gives a complete characterisation of probabilistic programs with nondeterminism, and they are sufficient to express many performance-style properties.\ To further explore this capability, we set out the definitions below.\ Their elementary details can be found elsewhere \cite{GW86,CGP99}.

\noindent {\it \textbf{Definition 1:} (Sub-distribution)\ For any finite state space S, the set of sub-distributions over S is
\begin{eqnarray}
\hspace*{-1ex}\begin{array}[t]{l}
\overline{S} \Defs \quad \{\Delta: S \rightarrow [0, 1] \mid \sum \Delta \le 1\} \ ,
\end{array}\label{eq:distribution}
\end{eqnarray}
the set of functions from S into the closed interval of reals [0, 1] that sum to no more than one \footnote{Probabilities that sum to less than one represent aborting behaviours. We do not discuss such program behaviours here.}}.

\noindent {\it \textbf{Definition 2:} (Labeled Markov Decision Process)\ A tuple (S, $\hat{s}$, $\textsf{A}$, $\textsf{L}$), where S is as defined above, $\hat{s} \in S$ is the initial state, $\textsf{A} : S \rightarrow 2^{\overline{S}}$ is a transition function, and $\textsf{L} : S \rightarrow 2^{AP}$ is a labeling function which assigns to each state, a subset of the set of atomic propositions AP that are valid for that state.}

\noindent {\it \textbf{Definition 3:} (Path)\ A path in an MDP is a non-empty finite or infinite sequence of states $s_{0}\stackrel{\alpha_{0}}{\longrightarrow}$ $s_{1}\stackrel{\alpha_{1}}{\longrightarrow}$ ... where $\alpha_{i} \in  \textsf{A}(s_{i})$ \ and \ $\alpha_{i}(s_{i + 1}) > 0$ \ for all $s_{i} \in S$.}

\noindent {\it \textbf{Definition 4:} (Absorbing state)\ A state $s_{i} \in S$ is said to be absorbing if no transition leaves this state after resolving all the nondeterministic selections in the state {\it i.e.}, $s_{i}\stackrel{\alpha_{i} = 1}{\longrightarrow} s_{i}$, and $s_{i}\stackrel{\alpha_{i} = 0}{\longrightarrow} s_{j}$  whenever $\mathord{i \ne j}$.}

A probabilistic computation tree formalises the notion of a probabilistic distribution over execution traces required to give semantic interpretation to temporal properties.\ Each step on a path will have an associated probability (often probability one for standard steps in the computation) --- and the probabilities on those individual steps when multiplied together, determine probability for paths ending up in a particular absorbing state.\ Our interest lies in only using such probabilities masses for {\it finite} paths. \\
\noindent {\it \textbf{Definition 5:} (Endpoint of a distribution)\ Any absorbing state s over the distribution $\Delta$ is said to be at the endpoint of the distribution.}

\noindent {\it \textbf{Definition 6:} (Random variable)\ A random variable is a non-negative real-valued function over the state space in which our programs operate.}

\noindent {\it \textbf{Definition 7:} (Expected value) \ For any bounded random variable $\alpha$ in $S \rightarrow \mathbb{R}_{\ge}$ and distribution $\Delta \in \overline{S}$, the expected value of $\alpha$ over $\Delta$ is defined
\begin{eqnarray}
\hspace*{-1ex}\begin{array}[t]{l}
\mathord{\displaystyle\int_{\Delta}\alpha} \Defs \quad \mathord{\displaystyle\sum (\alpha.s\ast\Delta.s)} \ ,
\end{array}\label{eq:expected}
\end{eqnarray}
for any state $s$ in the endpoint of the distribution $\Delta$. }

\subsection{The PCHOICE Operator}
In \cite{HJRMM03} Hoang T.S. {\it et al.} introduced a {\bf PCHOICE} operator in the standard AMN's operations --- similar to the probabilistic choice operator of Fig. \ref{fig:semantics}, which also permits the specification of probabilistic behaviours in a typical machine.\ This extension, captured in the the probabilistic Abstract Machine Notation (pAMN), and expressed in the pB method, describes probabilistic machines with an additional EXPECTATIONS clause\footnote{The complete framework encapsulates state variables and the operations on the states by the use of `clauses'.}.\ Ideally, probabilistic invariant properties are then defined as random variables over the machine's state, and encoded in the EXPECTATIONS clause.\ An invariant of this form is then an ``expected value-invariant''.\ Later on, we show how the pAMN can be used to specify abstract probabilistic system behaviours.\ A comprehensive list of the pAMN clauses can be found in \cite{Hoang05}.

\subsection{The EXPECTATIONS clause}
It gives a random variable $\xi$ over the program state, denoting the expected value-invariant, and an initial expression {\it e} which is evaluated over the program variables when the machine is initialised.\ The idea is that after arbitrary executions of the program, the expected value of $\xi$ at any given program state, is always at least the value of {\it e} initially \cite{HJRMM03}.

More formally, suppose a probabilistic machine has initialisation {\it INIT} and two operations {\it OpX} \ and {\it OpY} respectively, therefore, satisfying the probabilistic proof obligation for some expected value-invariant $\xi$ and initial expression $e$ would operationally (Fig. \ref{fig:semantics}) imply that\footnote{For random variables $R$, $R^{'}$, the implication-like relation $R \Rrightarrow R^{'}$ means $R$ is everywhere less than or equal to $R^{'}$.}
\begin{eqnarray}
\begin{array}{l}
\hspace*{-1ex}\begin{array}[t]{l}
 \quad  \xi \Rrightarrow \wP{{OpX}}{\xi}\quad \quad $and$ \quad \quad \xi \Rrightarrow \wP{{OpY}}{\xi} \quad $provided$ \quad e \ \Rrightarrow \wP{{INIT}}{\xi},
\end{array}
\end{array}\label{eq:probInv}
\end{eqnarray}
which then assures that\vspace{-.5cm}
\begin{eqnarray}
\begin{array}{l}
\hspace*{-1ex}\begin{array}[t]{l}
\quad e \ \Rrightarrow \mbox{$wp.{INIT}.{(wP.{({\mathrm{It}} \ OpX \ \sqcap \ OpY \ \mathrm{tI})}.{\xi})}$}.
\end{array}
\end{array}\label{eq:probdOP}
\end{eqnarray}
\noindent 

The operational interpretation of (\ref{eq:probdOP}) is that arbitrary interleaving of the operations {\it OpX} and {\it OpY} after the initialisation {\it INIT} should always result in a distribution over the final states (of variables) such that the expected value (with respect to invariant $\xi$) is at least the initial value specified by the expression $e$.\ Clearly, the conditions in (\ref{eq:probInv}) imply the operational interpretation of (\ref{eq:probdOP}).\ However, if there is a particular interleaving of the machine which demonstrates the failure of (\ref{eq:probdOP}), then it must be true that (\ref{eq:probInv}) has hitherto failed as well.\ The example below illustrates our argument.

\subsubsection{Example: A Simple Demonic Machine}
Fig. \ref{fig:simple} shows a pAMN (adapted from \cite{HJRMM03}) that captures the operations of a simple pB machine called Demon, with a single variable {\it cc}; the INVARIANT clause specifies that {\it cc} must be Integer-valued --- pB's prover always checks that this statement is true using the operational reasoning established in the previous section.\ Initially, {\it cc} is set to 0; the OPERATIONS clause contains operations {\it OpX} and {\it OpY}.\ {\it OpX} can either increment {\it cc} by 1 or decrement it by the same value both with probability $1/2$, while {\it OpY} just re-initialises {\it cc} to 0.\ The variable {\it nn} in the operation {\it OpX} is an output parameter which need not occur in the VARIABLES clause\footnote{It must however follow a similar machine declaration as $cc$ to enable its PRISM transformation.}.\ The EXPECTATIONS clause specifies the expected value-invariant $\xi$ to be the random variable {\it cc}, and the initial expression {\it e} to be 0, so that ``the expected value of {\it cc} over any endpoint distribution is never decreased below 0'' by the Demon's {\it OpX} and {\it OpY} operations.

\begin{figure}
\begin{center}
\scriptsize{
\begin{tabular}{l}
\hline\\
\mbox{\bf MACHINE} \ \ \ \quad \quad \quad \quad Demon \\
\mbox{{\bf SEES}\quad \quad \quad \quad \quad\ \quad \quad \ Int\_TYPE, Real\_TYPE}\\
\mbox {{\bf VARIABLES} \quad \quad \quad \quad $cc$}\\
\mbox{{\bf INVARIANT} \quad \quad \quad \quad $cc : INT$} \\
\mbox{{\bf EXPECTATIONS}\quad \quad $ \ real (0) \ \Rrightarrow \ cc$} \\
\mbox{{\bf INITIALISATION}\quad \quad $cc := 0$} \\
\mbox{{\bf OPERATIONS}}\\
\mbox{\quad  nn {\bf $\leftarrow$} \ OpX \quad \ = \quad \quad {\bf BEGIN}} \\
\hspace{1.5cm} \mbox{{\bf \ \ \ \ \quad \quad \ \quad \ PCHOICE} \ $frac$ (1, 2) \ {\bf OF} \ $cc : = cc + 1$} \\
\hspace{1.5cm} \mbox{{\bf \ \ \ \ \quad \quad \ \quad \ OR} \ $cc : = cc - 1$ \ {\bf END} $||$ $nn := cc$} \\
\ \hspace{1cm} \mbox{{OpY} \quad = \quad \quad $cc := 0$} \\
\mbox{{\bf END}} \\\\
\hline
\end{tabular}

\noindent
}
\caption{\rm A pB model specified in the pAMN.}\label{fig:simple}\vspace{-.75cm}
\end{center}
\end{figure}

In this example, the machine {\it fails} to satisfy the probabilistic proof obligation specified in (\ref{eq:probdOP}), and the reason is that
\begin{eqnarray}
\hspace*{-1ex}\begin{array}[t]{l}
\quad (\exists cc \in INT: \ \neg(cc \ \Rrightarrow \ \wP{OpY}{cc})). 
\end{array}\label{eq:expectedvalue}
\end{eqnarray}
The immediate expression captures the failure of the pB prover to establish the proof obligations in (\ref{eq:probInv}).\ However, in terms of a distribution viewpoint, it is possible to see exactly why this failure of the pB prover would similarly  result in the failure of (\ref{eq:probdOP}).\ Consider the program fragment
\begin{eqnarray}
\begin{array}{l}
 INIT; OpX; (\ccc{OpY}{(nn \ge 0)}{{\mathrm{skip}}});
\end{array}\label{eq:probinit}
\end{eqnarray}
it yields the distribution given by
\[ \delta = \left\{ \begin{array}{l}
 \hspace*{-1ex}\begin{array}[t]{l}
\ cc := 0 \ \ \ \quad @ 1/2 \\
\ cc := -1 \quad @ 1/2
\end{array}
\end{array}\right.\]
over the final state of the random variable {\it cc}.\ Calculating the expected value over this distribution we get
\begin{eqnarray}\nonumber
\hspace*{-1ex}\begin{array}[t]{l}
\mathord{\displaystyle\int_{\delta} cc} = 1/2 \times 0 + 1/2 \times -1 \quad \equiv  - 1/2, 
\end{array}
\end{eqnarray}
which is clearly a violation of the conditions specified in the EXPECTATIONS clause.

In this simple case, it is clear that the failure to establish the pB proof obligation corresponds to an exact result distribution over endpoints that demonstrates the failure.\ Currently pB provers do not provide diagnostic information necessary to give practitioners the much needed operational intuition (in terms of distributions) for locating failures.\ This becomes even more complicated with increasing size and complexity of the pB machines.\ For such cases, we simply rely on the model checking capabilities of state-of-the-art tools like PRISM. To do this, we need to translate the pB machines to their equivalent PRISM models, and use the latter's algorithmic analysis to attempt to locate failures.

In the sections that follow, we show how the analysis of an equivalent PRISM representation of a pB machine can provide a link between the operational viewpoints (distribution-centered) of both a proof-based environment (encapsulating probabilistic invariants), and a model checking platform.\ Such a link is key to getting a better understanding of the expected value-invariants over endpoint probabilistic distributions.

\section{PRISM Reward Specification}
The PRISM model checker permits models to be augmented with information about rewards (or costs).\ The tool can analyse properties which relate to the expected values of the rewards.\ A reward structure \cite{KNP07} can be used to represent additional information about the system the MDP (the model types in this paper) represents --- for example, the expected number of packets sent (or lost) on a protocols request.

The temporal logic probabilistic CTL \cite{HJ94} has been extended in \cite{AHK03} to allow for reward-based specifications as constraints of the type that express {\it reachability}, {\it cumulative} and {\it instantaneous} rewards to model checkers.\ However, for the purpose of this work, the instantaneous variant is useful.

\noindent {\it \textbf{Definition 8:} (Expected instantaneous reward)\ The probabilistic CTL permits reward properties of the form \ $\mathord{{R}_{\sim r} [\mathbb{I}^{= k}]}$, at time-step $k$, where \ $\mathord{\sim \in \{<, \le, \ge, >\}}$, $\mathord{r \in \mathbb{R}_{\ge}}$ and $\mathord{k \in \mathbb{N}}$.\ The reward formula $\mathord{{R}_{\sim r} [\mathbb{I}^{= k}]}$\footnote{For MDP's we require Rmin or Rmax; this is allowed in PRISM by enabling the sparse engine in the tool's options menu.}is true if from some initial state $s_{0}$, the expected state reward at time-step $k$ meets the bound $\sim r$.\ For example the specification $\mathord{R_{\ge 50}[\mathbb{I}^{=2}]}$ could be interpreted to mean that the expected number of packets sent by the protocol after two time-steps is at least 50.}

\section{Quantitative Safety and pB Machines}\label{sec:safety} 
Formal approaches necessitate that every safe system be {\it invariant-driven} \cite{Dijkstra76}.\ Therefore, quantitative safety properties can be proved by verifying invariants.\ The EXPECTATIONS clause of a pB machine encapsulates the machine's safety property.\ However, an interesting dimension in investigating pB safety properties is whether it is possible to find a nondeterministic selection $\sqcap_{0\le n \le N} \ P_{n}$ of the operations of the machine that demonstrates the failure of the invariants --- this schedule must then lead to the problem of locating counterexamples for probabilistic model checking \cite{HK07}.

In \cite{MMG08} McIver {\it et al.} set out a strategy for computing a ``refutation-of-safety'' certificate for probabilistic systems using model checking techniques.\ The existence of a certificate corresponds to an invariant failure in our own context of safety.\ In the next section, we present an automation which demonstrates the key features of that strategy and in particular show how it can be used to investigate pB safety properties.\ The theorem below is fundamental for investigating safety properties specified for pB machines.

\noindent {\it \textbf{Theorem:}\ Given a pB machine invariant $\xi$ and initial expression {\it e} encapsulated in the machine's EXPECTATIONS clause, let $\Delta$ be the finite result distribution over endpoints for the interleaving  $\mathord{\mathrm{It} \ \sqcap_{0\le n \le N} \ P_{n} \ \mathrm{tI}}$ after the machine's initialisation INIT.\ A safe machine always guarantees that
\begin{eqnarray}
\begin{array}{l}
e \ \Rrightarrow \wP{INIT}{(\wP{(\mathrm{It} \ \sqcap_{0\le n \le N} \ P_{n}\ \mathrm{tI})}{\xi}) \ \Rightarrow \quad \displaystyle\int_{{{\Delta}}} \xi} \ \ge \ e,
\end{array}\label{eqn:theorem}
\end{eqnarray}
\noindent provided the pB machine's prover will always discharge the proof obligations given by $\mathord{e \ \Rrightarrow \wP{INIT}{\xi}}$ \ and \ $\mathord{\xi \ \Rrightarrow \wP{(\mathrm{It} \ \sqcap_{0\le n \le N} \ P_{n} \ \mathrm{tI})}{\xi}}$
} respectively.

\noindent {\it \textbf{Corollary:}\ Suppose it is always possible to split the distribution $\Delta$ into finite sub-distributions $\delta_{k}$ for all $k \ge 0$ where $\delta_{k}$ is the result distribution on the $kth$ iteration, then
\begin{eqnarray}
\begin{array}{l}
\mathord{(\forall k \ge 0 \wedge \delta_{k} \in \Delta: \quad \displaystyle\int_{\delta_{k}} \xi} \ \ge \ e) \ .
\end{array}\label{eqn:corollary}
\end{eqnarray}}
The usefulness of the corollary is in the practical demonstration of the theorem.\ Its simplicity is captured by the fact that, if a probabilistic computation tree which interprets the execution traces of the machine is available to a verifier, then these traces must be finite with respect to the result distributions they represent.\ Using bounded model checking techniques, it then suffices to argue that only finite values of $k$ are required to establish the proof obligation for a safe pB machine.\ More so, with state-of-the-art probabilistic model checking tools such as PRISM, it is possible to identify a sub-space of the entire distribution in which the failure is located (if any).

\section{YAGA: A pAMN To PRISM Translator}\label{sec:YAGA}
In section \ref{sec:safety}, we stated a theorem that is central to defining safety features for pB machines with a finite trace distribution over  endpoints.\ The corollary of that theorem has the practical interpretation that, {bounded model checking techniques when equipped with reward structures} are likely to provide an intuition to locating invariant failures in their transformed proof-based models.\ In this section we discuss a language-level translator nicknamed \texttt{YAGA}\footnote{The name YAGA is coined from an Igbo (a language largely spoken in southeastern Nigeria) word  --- YAGAzie, which literally means ``may it go well ...''.\ In reality, it could be argued that YAGA is simply Yet Another Gangling Automation.} and with architectural framework shown in Fig. \ref{fig:YAGA}.\ \texttt{YAGA}, a java-based implementation of the algorithm {pAMN2PRISM} (shown in Appendix A) is a prototype system that essentially takes a pAMN framework, encapsulating a pB model (with syntactic checks supposedly discharged) as its input parameter, and generates its precise probabilistic action systems representation in the PRISM language.\ The associated reward structure of the generated PRISM model is inherited from the pB machine's EXPECTATIONS clause.\ The PRISM model checker then readily offers its temporal logic specification (as probabilistic CTL formulas) which can easily be checked by conducting experiments on the transformed model.\ The experimental results are sufficient to validate (or refute) the probabilistic invariants specified in the abstract machine's EXPECTATIONS clause.

\subsection{Overview: YAGA Transformation Rules}
We summarise the transformation rules as follows.\ YAGA's algorithmic interpretation is in Appendix A.

\begin{figure}
\begin{center}\includegraphics[scale = 0.25]{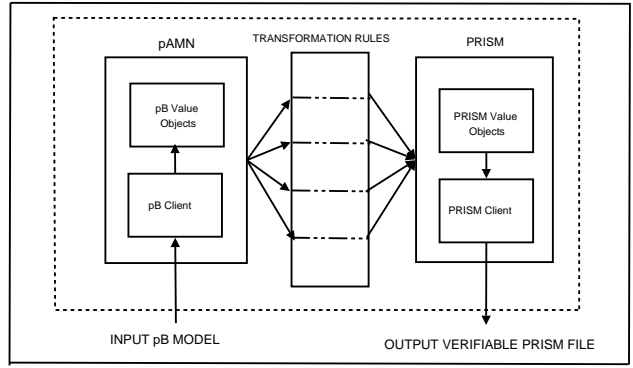} \end{center}\vspace{-.5cm}
\caption{\rm YAGA - Architectural Overview}\label{fig:YAGA}\vspace{-.5cm}
\end{figure}

\subsubsection{Main Module}
\noindent {\bf PRISM constants list:} Are constructed from the pB machine's parameter list (if any) and the CONSTANTS clause.\ The type of a constant is implicitly checked from the PROPERTIES clause.

\noindent{\bf PRISM formula list:} Are auto-generated as {\it atomic} predicates from the pB machine's PROPERTIES and INVARIANTS clauses.

\noindent{\bf PRISM module name:} Is the pB machine's name.

\noindent{\bf PRISM variables declaration and initial values list:} Are constructed from each variable in the VARIABLES clause, its type in the INVARIANT clause, and its initial values from the INITIALISATION clause.\ The lower and upper limits of the variables are respectively the default lowest values of their types, and a bound specified from the PRISM constants list (above).

\noindent{\bf PRISM update statements:} Each update statement is labeled with the operation's name from the pB machine.\ In addition,
\begin{itemize}
\item [(a)] its guard is inherited from the guard in the pB machine's OPERATIONS clause and strengthened by the formulas in the PRISM formula list, such that
\item [(b)] the choice of a selection of formula is dependent on the expressions in the pB machine's update statement.\ For each update, YAGA checks that the formula-dependent expressions are included in the PRISM guard.
\end{itemize}

\subsubsection{Counter Module}
The counter module is a vital encoding that helps enumerate the distributions in $\delta_{k}$ (in corollary) for finite $k$ steps.\ To capture this behaviour in a model checking environment, we apply the following rules.

\noindent{\bf PRISM module name:} Counter.

\noindent{\bf PRISM variable declaration and initial value:} Variable {\it count} is initially set to 0 and bounded by $\mathord{(MAX\_COUNT + 1)}$.

\noindent{\bf PRISM update statement:} Each update is constructed to synchronise with the updates in the main module.\ They can only increment the {\it count} variable by 1 on each action.\ In addition, this module also contains a similar unsynchronised update statement which ensures we will eventually reach $\mathord{(MAX\_COUNT + 1)}$.

\subsubsection{Reward Structure}
The specific reward structure is inherited from the pB machine's EXPECTATIONS clause --- states where the {\it count} variable equals $\mathord{(MAX\_COUNT + 1)}$ are worth the random variable value specified in the EXPECTATIONS clause plus $\mathord{(MAX\_COUNT)}$\footnote{This padding is to ensure the PRISM engine is consistent with computing positive instantaneous rewards.\ Finally, we subtract this parameter from the PRISM computed reward value.}.

\ However, to make the construction of our reward structures precise for model checking, we note more formally as a result of the theorem in sec. \ref{sec:safety} that

\noindent  {\it \textbf{Remark 1:} Given any pB machine invariant $\xi$ and initial expression {\it e}, then from an initial state $s_{0}$, after the machine's initialisation INIT, any arbitrary interleaving $\mathord{\mathrm{It} \ \sqcap_{0\le n \le N} \ P_{n} \ \mathrm{tI}}$ must guarantee that:
\begin{eqnarray}
\begin{array}{l}
(k:\in[0, MAX\_COUNT + 1]: e \Rrightarrow \wP{INIT}({\wP{(\mathrm{It} \ \sqcap_{0\le n \le N} P_{n} \ \mathrm{tI})}{\xi}}) \Rightarrow
\hspace*{-1ex}\begin{array}[t]{l}
Rmin_{=?} [\mathbb{I}^{= k}] \ge \xi.s_{0})
\end{array}
\end{array}\label{eq:bdmodelchecking}
\end{eqnarray}
such that the expected minimum instantaneous reward at the $kth$ step is worth the random variable value of the EXPECTATIONS clause plus $\mathord{MAX\_COUNT}$.\ We recall that the {\it Counter} module keeps explicit track of the $kth$ time-step, and the expected value here is captured by the sub-distribution in (\ref{eqn:corollary}).
}\\
\noindent {\it \textbf{Remark 2:} However, if there exists some {\it k} such that (\ref{eq:bdmodelchecking}) above fails to hold, then $\xi$ cannot be an invariant}.

\section{Case Study: A Library Bookkeeping System}\label{sec:casestudy}
We present a pB machine which captures the basic operations underlying the accounting package of a library system --- the implication of an undischargeable proof obligation of the machine on the performance of the library was an open problem in \cite{HJRMM03}.\ The state of the machine contains four variables: {\it booksInLibrary}, {\it loansStarted}, {\it loansEnded} and {\it booksLost} which are respectively used to keep track of: the number of books in the library, the number of book loans initiated by the library, the number of book loans completed by the library, and the number of books possibly never returned to the library.

Initially, the machine has two operations: {\it StartLoan}, to initiate a loan on a book, and {\it EndLoan}, to terminate the loan of a book.\ The {\it StartLoan} operation has a precondition that there are books available for loan; it decrements {\it booksInLibrary} and increments {\it loansStarted}; when a book is returned, the {\it EndLoan} operation reverses the effect of the {\it StartLoan} operation by recording that either the book ``really is'' returned, or is actually reported lost with some probability $pp$, so that {\it booksLost} is incremented.

The machine uses the random variables {\it loansEnded} and {\it booksLost} to record the expected losses of the number of books over time.\ Since with a probability $pp$ a book is lost on each {\it EndLoan} operation, the library system would then be expected to lose a proportion $pp$ over a number of {\it EndLoan} operations.\ However, to ensure that the library is always in the business of books lending, we define the expected value-invariant
\begin{eqnarray}
\hspace*{-1ex}\begin{array}[t]{l}
\displaystyle\int_{\Delta} \ (pp \times loansEnded - booksLost) \ \ge \ 0,
\end{array}\label{eq:inv}
\end{eqnarray}
\noindent which captures the idea that the expected value of the random variable $\mathord{pp \times loansEnded - booksLost}$
over its endpoint distributions $\Delta$ can never be decreased below 0.\ This is indeed a safety property for the library and ought
to be checked throughout its lifetime to ensure it is not violated by its future designs.\ Below, we present two designs of the library and also keep in mind the property specified in (\ref{eq:inv}).

\subsection{A Safe Library Bookkeeping System}
Since there are no restrictions on when the operations of the machine can be invoked, except for the obvious preconditions on {\it StartLoan} and {\it EndLoan}; the specification of a safe library system is the nondeterministic choice given by
\begin{eqnarray}
\hspace*{-1ex}\begin{array}[t]{l}
SafeLibrary \Defs \ \mathrm{It} \ StartLoan \ \sqcap \ Endloan \ \mathrm{tI} \ .
\end{array}\label{eq:spec1}
\end{eqnarray}

\subsection{An Unsafe Library Bookkeeping System}

Suppose that to enable the library accountant do a periodic stock take of the library transactions, a new operation called {\it StockTake} is introduced into the system.\ The {\it StockTake} operation is very similar to the initialisation, except for an extra output ({\it totalCost}) to record the cost of replacing the books lost up to the time of doing a stock take.\ Again, we augment (\ref{eq:spec1}) and give another specification of the library as
\begin{eqnarray}
\hspace*{-1ex}\begin{array}[t]{l}
UnsafeLibrary \Defs \ \mathrm{It} \ StartLoan \ \sqcap \ Endloan \ \sqcap \ StockTake \ \mathrm{tI} \ .
\end{array}\label{eq:spec2}
\end{eqnarray}
The complete pB machine describing the library (all of its three operations) is shown in Appendix A.
\begin{figure}\vspace{-3.5cm}
\begin{center}\includegraphics[scale = 0.45]{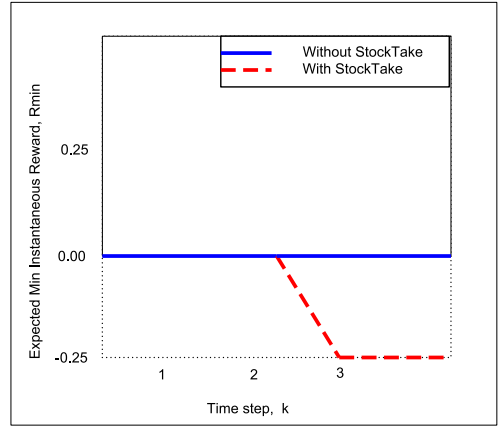}\end{center}\vspace{-5cm}
\caption{\rm Library Bookkeeping System}\label{fig:result}\vspace{-.5cm}
\end{figure}

\section{PRISM Experimental Results}
In this section we report experimental results that are indeed performance-style characterisations of the two designs of our library model --- the safe library (\ref{eq:spec1}) and the unsafe library (\ref{eq:spec2}).\ Our interest lies in justifying the reason why one design of the library (without stockTake) is safe and why the other (with stockTake) is unsafe.\ We note that our safety property of concern is captured by (\ref{eq:inv}).

To enable us carry out this performance analysis, we quickly use the capabilities of YAGA.\ The equivalent PRISM representation of the pB machine discussed in the previous section as generated by YAGA is shown in Appendix A.\ From {Remark 1}, our obvious reward specification becomes
\begin{eqnarray}
\hspace*{-1ex}\begin{array}[t]{l}
(\forall \ 0 \le k \le MAX\_COUNT:   \quad Rmin_{=?} [\mathbb{I}^{= k}] - MAX\_COUNT)\ . 
\end{array}\label{eq:spec3}
\end{eqnarray}
The requirement for a safe library system is that for all time-steps $k$, $Rmin$ is never decrease below zero.\ However, our experimental result (shown in Fig. \ref{fig:result}) reveals that indeed the unsafe library violates this safety property after just three time-steps ($k$ = 3) of its execution --- that is for $MAX\_COUNT$ = 2, $\mathord{pp = 0.5}$, and $totalBooks = 1$, the expected minimum instantaneous  reward $Rmin$ of the unsafe library is -0.25.\

A quick conclusion that can be drawn from our analysis is that introducing the  ``demonic'' StockTake operation has the adverse effect of subverting the overall performance outlook of the library system.\ Our result is a practical demonstration of the claim by Hoang T.S. {\it et al.} \cite{HJRMM03} in an attempt to explain why the presence of the StockTake operation would result in a failure of the proof obligation for the probabilistic invariant property in (\ref{eq:inv}).\ In the proof-based system, reaching this conclusion was practically impossible.

\section{Conclusion and Future Work}
This paper has explored the practical application of reward-based specifications of bounded model checking techniques \cite{CBC03} to locate failures in the context of proof-based verification for simple safety properties of probabilistic systems.\ We demonstrated the rich benefits that can be derived by complementing proof-based probabilistic verification techniques with a model checking performance-style evaluation, in a manner that has never been previously explored.

Our contribution is seen as a first attempt at fully integrating quantitative performance analysis to systems design at early stages of development.\ Our method scales in this regard since it can be carried out at the level of source code  and hence can guide system developers with decisions on a choice of design most suitable for implementation.

However, in other to fully integrate this performance-style analysis into software development process, we intend to, in the future,
incorporate a diagnostic mechanism based on counterexamples location employed in \cite{HK07,HS07} to YAGA.\ Our intention is that such a mechanism will report explicit cause(s) of failure by also exploring the backward analysis strategy in \cite{MMG08}.\ It is our belief that these enhancements would provide a useful performance analysis suite for probabilistic systems developed in the pB language.

\section{Acknowledgement}
The author is grateful to his thesis supervisor, A.K. McIver for her unflinching research guidance.\ We also wish to express our gratitude to Dave Parker of the Oxford University computing laboratory for his technical input on matters relating to the PRISM tool.\ Finally, to the anonymous reviewers whose comments have been invaluable in improving the preliminary draft of this paper, we say a big thank you.

\bibliographystyle{eptcs} 

\begin{figure}
\scriptsize{
$
\begin{array}{l}
 \mbox{\bf \large{Appendix A}}\\\\
\hline
\noindent
\mbox{{\bf MACHINE} \ ProbabilisticLibrary (totalBooks, cost)}\\
\mbox{{\bf SEES}\ Real\_type}\\
\mbox{{\bf CONSTANTS} \ pp}\\
\mbox{{\bf PROPERTIES}\ $pp \in REAL \ \wedge \ pp \le real \ (1) \ \wedge \ real \ (0) \le pp$}\\
\mbox {{\bf VARIABLES} \ {\it booksInLibrary, loansStarted, loansEnded, booksLost, totalCost}}\\
\mbox{{\bf INVARIANT} \ {\it booksInLibrray, loansStarted, loansEnded, booksLost, totalCost\ $\in$ \ NATURAL \ $\wedge$ \ loansEnded $\le$ loansStarted}}\\
\mbox{ \quad \quad  \quad  \quad \quad \ \     $\wedge$\ {\it booksInLibrary + booksLost + loansStarted - loansEnded = totalBooks}}\\
\mbox{{\bf EXPECTATIONS}\ $real \ (0) \Rrightarrow pp \times real \ (loansEnded) - real (booksLost)$}\\
\mbox{{\bf INITIALISATION} {\it booksInLibrary, loansStarted, loansEnded, booksLost, totalCost := totalBooks, 0, 0, 0, 0 }}\\
\mbox{{\bf OPERATIONS} \ \ \ StartLoan \quad \ = \quad {\bf PRE} {\it booksInLibrary $>$ 0} {\bf THEN}}\\
\mbox{   \quad \quad  \quad  \quad \quad    \quad  \quad \quad \quad \quad \quad \quad   {\it \ \quad \quad \ booksInLibrary := booksInLibrary - 1 $||$ loansStarted := loansStarted + 1 {\bf END}};}\\
\mbox{  \quad \quad  \quad  \quad \quad    \quad  \quad       \ \ EndLoan \quad \ = \quad {\bf  PRE} {\it loansEnded $<$ loansStarted} {\bf THEN}}\\
\mbox{ \quad \quad  \quad  \quad \quad    \quad  \quad \quad \quad \quad \quad \quad      {\bf \ \quad \quad \ PCHOICE} {\it pp} {\bf OF} {\it booksLost := booksLost + 1}}\\
\mbox{ \quad \quad  \quad  \quad \quad    \quad  \quad \quad \quad \quad \quad \quad      {\bf \ \quad \quad \ OR} {\it booksInLibrary := booksInLibrary + 1} {\bf END} $||$ {\it loansEnded : = loansEnded + 1} {\bf END}};\\
\mbox{\quad  \hspace{1.7cm} { StockTake} \quad = \quad {\bf BEGIN} {\it totalCost := cost $\times$ booksLost $||$ booksInLibrary := booksInLibrary + booksLost $||$}}\\
\mbox{   \quad \quad  \quad  \quad \quad    \quad  \quad  \quad \quad  \quad  \quad  \quad \quad \quad \ {\it \ loansStarted := loansStarted - loansEnded $||$ loansEnded := 0 $||$ booksLost := 0 \quad {\bf END \quad \quad \ \ }}} \\
\mbox{{\bf END}}\\
\hline

\end{array}
$
\noindent
}
\caption{\rm The pB Model of Section (\ref{sec:casestudy})}\label{fig:pB}
\end{figure}

\begin{figure}
\scriptsize{
$
\begin{array}{l}
\hline
\noindent
\mbox{{\bf const} \ {\it totalBooks;}}\\
\mbox{{\bf const} \ {\it cost;}}\\
\mbox{{\bf const \ double} \ {\it pp;}}\\
\mbox{{\bf const} \ {\it MAX\_COUNT;}}\\
\\
\mbox{{\bf formula} \ {\it formula0 \quad = \quad  (loansEnded $\le$ loansStarted);}}\\
\mbox{{\bf formula} \ {\it formula1 \quad = \quad (booksInLibrary + booksLost + loansStarted - loansEnded = totalBooks);}}\\
\mbox{{\bf formula} \ {\it formula2 \quad = \quad (pp $\le$ 1);}}\\
\mbox{{\bf formula} \ {\it formula3 \quad = \quad (0 $\le$ pp);}}\\
\\
\mbox{{\bf module} \ {ProbabilisticLibrary}}\\
\mbox{{\it booksLost:[0..totalBooks]}	\hspace{2cm} {\bf init}	\  $0$;}\\
\mbox{{\it totalCost:[0..totalBooks]}	\hspace{2.1cm} {\bf init}	\  $0$;}\\
\mbox{{\it loansEnded:[0..totalBooks]}	\hspace{1.8cm} {\bf init}	\  $0$;}\\
\mbox{{\it loansStarted:[0..totalBooks]}	\hspace{1.7cm} {\bf init}	\  $0$;}\\
\mbox{{\it booksInLibrary:[0..totalBooks]}	\hspace{1.4cm} {\bf init}	\  $totalBooks$;}\\
\\
\mbox{{\it [StockTake] \ formula1 \& formula0 $\rightarrow$ (totalCost $'$ = cost * booksLost) \& (booksInLibrary $'$ = booksInLibrary + booksLost)}}\\
\mbox{{\it \hspace{4.4cm}\& (loansStarted $'$ = loansStarted - loansEnded) \& (loansEnded $'$ = 0) \& (booksLost $'$ = 0);}}\\
\mbox{{\it [StartLoan]\ (booksInLibrary $>$ 0) \& formula1 \& formula0 \& (loansStarted + 1 $\le$ totalBooks)\ $\rightarrow$}}\\
\mbox{{\it \hspace{4.4cm}  (booksInLibrary $'$ = booksInLibrary - 1) \& (loansStarted $'$ = loansStarted + 1);}}\\
\mbox{{\it [EndLoan]\ (loansEnded $<$ loansStarted) \& formula2 \& formula1 \& formula3 \& formula0 \ $\rightarrow$}}\\
\mbox{{\it \hspace{4.4cm}  pp: (booksLost $'$ = booksLost + 1) \& (loansEnded $'$ = loansEnded + 1)}}\\
\mbox{{\it \hspace{4.4cm} + (1 - pp): (booksInLibrary $'$ = booksInLibrary + 1) \& (loansEnded $'$ = loansEnded + 1);}}\\
\mbox{{\bf endmodule} }\\
\\
\mbox{{\bf label} \ {\it ``expectations''	=	(pp *  loansEnded -  booksLost $\ge$  0);}}\\
\\
\mbox{{\bf module} \ {Counter}}\\
\mbox{{\it count:[0..MAX\_COUNT + 1]}	\quad {\bf init}	\  $0$;}\\
\\
\mbox{{\it [StockTake]\ (count + 1 $\le$ MAX\_COUNT + 1) $\rightarrow$ (count $'$ = count + 1);}}\\
\mbox{{\it [StartLoan]\ (count + 1 $\le$ MAX\_COUNT + 1) $\rightarrow$ (count $'$ = count + 1);}}\\
\mbox{{\it [EndLoan]\ (count + 1 $\le$ MAX\_COUNT + 1) $\rightarrow$ (count $'$ = count + 1);}}\\
\mbox{{\it []\ (count + 1 $\le$ MAX\_COUNT + 1) $\rightarrow$ (count $'$ = count + 1);}}\\

\mbox{{\bf endmodule} }\\
\\
\mbox{{\bf rewards} }\\
\mbox{{\it (count = MAX\_COUNT + 1) :\quad (pp *  loansEnded -  booksLost) + MAX\_COUNT;}}\\

\mbox{{\bf endrewards} }\\

\hline

\end{array}
$
\noindent
}
\caption{\rm A YAGA-Generated PRISM Representation of Fig. (\ref{fig:pB})}\label{fig:PRISM}
\end{figure}

\begin{figure}
\begin{center}
$
\normalsize
\begin{array}{l}
\hline
${Algorithm pAMN2PRISM (pB\_model\_in\_pAMN)}$\\
\hline\\
\mbox{Required: An interface for pB syntax, PRISM syntax, and regular math operators}.\\
\mbox{Reserved: MAX\_COUNT (constant), count (variable)}\\
\mbox{1: \quad get pAMN parameter list if any}   \\
\mbox{2: \quad create a map object with pAMN clauses as the the keys.\ Insert their respective values}\\
\mbox{3: \quad construct value objects with (1) and (2)}\\
\mbox{4: \quad set module type as MDP}\\
\mbox{5: \quad construct PRISM constants list from the pAMN parameter list and PROPERTIES key}\\
\mbox{6: \quad construct PRISM formula list as atomic predicates from the INVARIANT and}\\
\mbox{\quad \quad PROPERTIES keys}\\
\mbox{7: \quad get PRISM module name from MACHINE key (declare module name)}\\
\mbox{8: \quad construct PRISM variables list and their initial values from the VARIABLES}\\
\mbox{\quad \quad and INVARIANT keys and the pAMN parameter list (if any)}\\
\mbox{9: \quad {\bf for} the OPERATIONS key in the map object} \ {\bf do}\\
\mbox{\quad\quad \quad get the list of operations}\\
\mbox{\quad\quad \quad {\bf for} each operation in the list {\bf do}}\\
\mbox{\quad\quad \quad \quad {\bf for} each guard and update statement in operation {\bf do}}\\
\mbox{\quad\quad \quad \quad \quad check variables dependency on (6)} \\
\mbox{10: \quad construct PRISM update statements with the pAMN operations names as the }\\
\mbox {\quad \quad \ \ action labels}\\
\mbox{11: \quad declare {\bf endmodule}}\\
\mbox{12: \quad construct expectations label from the EXPECTATIONS key in map object}\\
\mbox{13: \quad declare {\bf module} counter}\\
\mbox{14: \quad declare count variable, initialised to zero and bounded by MAX\_COUNT + 1}\\
\mbox{15: \quad {\bf for} the OPERATIONS key in the map object} \ {\bf do}\\
\mbox{\quad\quad \quad get the list of operations}\\
\mbox{\quad\quad \quad {\bf for} each operation in the list {\bf do}}\\
\mbox{\quad\quad \quad \quad construct a synchronised update statement (increment) on the count}\\
\mbox{\quad\quad \quad \quad  variable with (10)}\\
\mbox{16: \quad {construct an unsynchronised update statement on the count variable}}\\
\mbox{17: \quad declare {\bf endmodule}}\\
\mbox{18: \quad declare PRISM {\bf rewards}}\\
\mbox{19: \quad for states where (count = MAX\_COUNT + 1)} \\
\mbox{20: \quad {set} reward to random variable value on EXPECTATIONS key plus MAX\_COUNT} \\
\mbox{21: \quad declare {\bf endrewards}}\\\\
\hline\\

\end{array}
$

\end{center}

\caption{\rm An Algorithmic Description of YAGA}
\end{figure}

\end{document}